\documentclass[aps,prd,twocolumn,superscriptaddress,showpacs]{revtex4}
\usepackage{graphics}
\usepackage{epstopdf}
\usepackage{graphicx}
\usepackage{dcolumn}
\usepackage{bm}
\usepackage{amsmath}
\usepackage{color}
\usepackage{ulem}

\newcommand{\be}{\begin{equation}}
\newcommand{\ee}{\end{equation}}
\newcommand{\beq}{\begin{eqnarray}}
\newcommand{\eeq}{\end{eqnarray}}

\def \gta {\mathrel{\vcenter{\hbox{$>$}\nointerlineskip\hbox{$\sim$}}}}

\def\t13{\mathrel{{\theta_{13}}}}
\def\y12{\mathrel{{\tan^2 \theta_{12}}}}
\def\c2{\mathrel{{\chi^2 }}}

\newcommand{\n}{neutrino}
\newcommand{\ns}{neutrinos}

\newcommand{\fb}{FB}
\newcommand{\fl}{\textit{Fermi}-LAT}
\newcommand{\ic}{IceCube}
\newcommand{\hw}{HAWC}

\begin{document}


\title{A multi-messenger study of the Fermi Bubbles: very high energy gamma rays and neutrinos}

\author{Cecilia Lunardini} \email{Cecilia.Lunardini@asu.edu}
\affiliation{Department of Physics, Arizona State University, Tempe,
  AZ 85287-1504} 

\author{Soebur Razzaque} \email{srazzaque@uj.ac.za}
\affiliation{Department of Physics, University of Johannesburg, PO Box
  524, Auckland Park 2006, South Africa}
 
\author{Lili Yang} \email{lili.yang@ung.si} \affiliation{Laboratory
  for Astroparticle Physics, University of Nova Gorica, Vipavska 13,
  5000 Nova Gorica, Slovenia}


\begin{abstract} 
The Fermi Bubbles have been imaged in sub-TeV gamma rays at Fermi-LAT, and, if their origin is hadronic, they might have been seen with low statistics in $\sim 0.1- 1$ PeV neutrinos at IceCube.  
We discuss the detectability of these objects at the new  High Altitude Water Cherenkov (HAWC) gamma ray detector. \hw\ will view the North Bubble for $\sim 2-3$ hours a day, and will map its spectrum at 0.1--100 TeV. 
For the hard primary proton spectrum required to explain five events at \ic, a high significance detection at \hw\ will be achieved in less than 30 days.  The combination of results at \hw\ and \ic\ will substantiate the hadronic model, or constrain its spectral parameters. 
\end{abstract}
 
\maketitle

{\it Introduction.\ ---} 
Studying the sky at Very High Energies (VHE) of $\sim$ 100 GeV and beyond is at the frontier of astrophysics today. Until very recently, Gamma Ray Astronomy has been the only avenue to probe the VHE sky, and has revealed the existence of powerful natural particle accelerators, whose physics is still largely unknown \cite{Aharonian:2008zza}.  It is expected that collisions of accelerated hadrons with the ambient medium are at least partly responsible for VHE gamma ray emission from these sources. The same processes also produce \ns, and therefore a VHE \n\ counterpart of the gamma ray flux is expected.  In 2013, the \ic\ Neutrino Observatory  reported the detection of astrophysical \ns\ in the 0.1--1 PeV energy range \cite{Aartsen:2013bka,Aartsen:2013jdh,Aartsen:2014gkd}, which opened a new avenue of multi-messenger exploration of the VHE sky.  The complementarity between  gamma rays and \ns\ is strong, mainly because  they both probe the sites of the cosmic-ray accelerators in the universe.

\begin{figure}[htbp]
\centering
\includegraphics[width=3.3in]{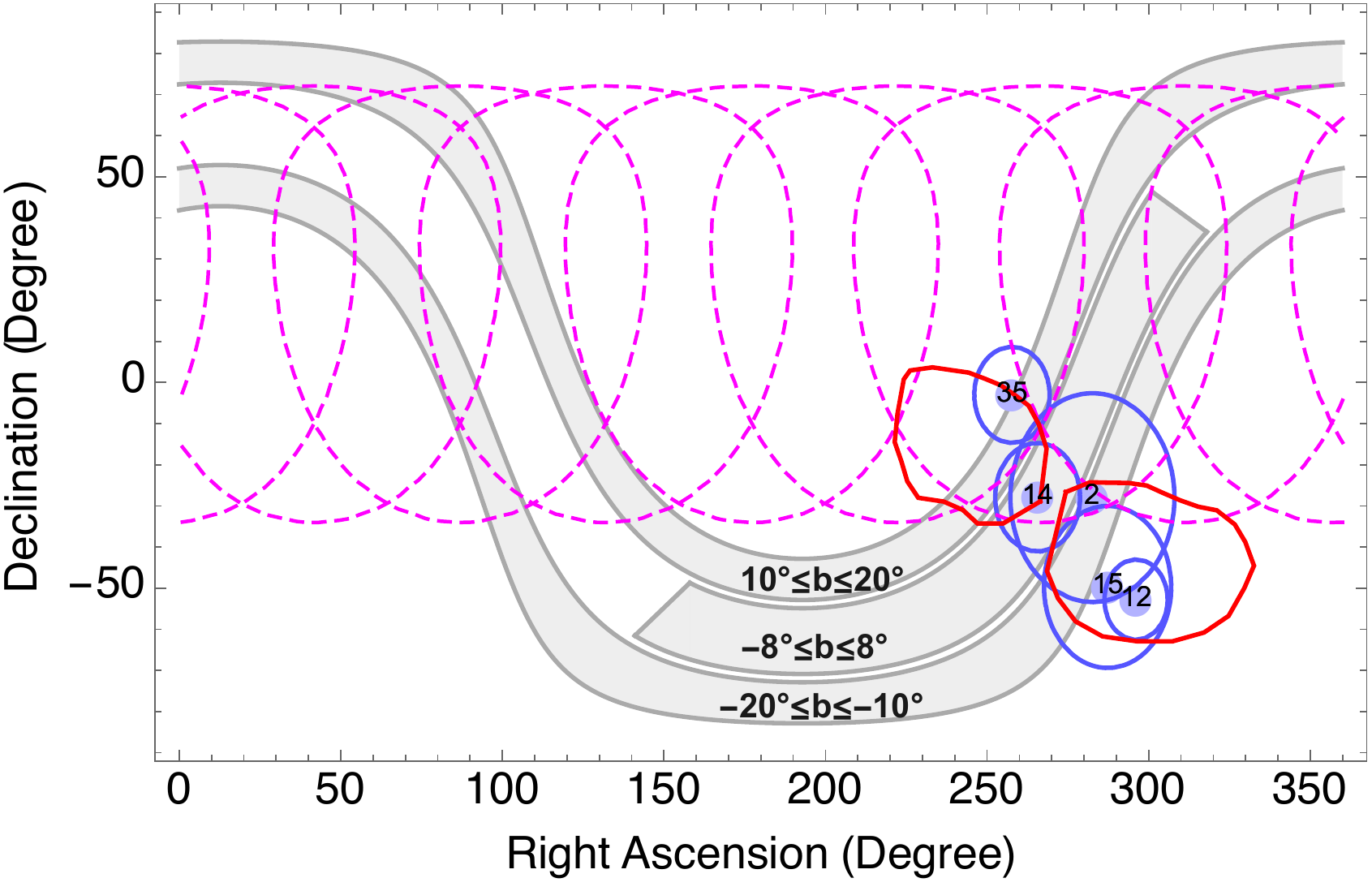}
\caption{Map of the Fermi Bubbles in equatorial coordinates (red solid contours). The blue dots show the \ic\ events that are spatially strongly correlated with the \fb\, with their sequence numbers \cite{Aartsen:2014gkd} and their positional errors (blue ellipses, see text).  Also shown are the contours of the HAWC's field of view ($0.6\leq \cos\theta\leq 1$), for which the effective area is published \cite{Zaborov:2012jg}, at several times of the day (magenta dashed contours). The shaded curves represent the inner Galaxy region (corresponding to Galactic coordinates $-80^\circ\le l\le 80^\circ$, $-8^\circ\le b\le 8^\circ$) and the low intermediate latitude region ($0^\circ\le l\le 360^\circ$, $10^\circ\le |b| \le 20^\circ$), where \fl\ has measured diffuse Galactic gamma-ray emission \cite{FermiLAT:2012aa}. }
\label{FBgal}
\end{figure}

Since the \n\ data are still limited by low statistics, a multi-messenger study of a single, specific source is possible only for the closest and most powerful emitters in the sky, like the Galactic Center and the Fermi Bubbles (\fb).  The \fb\ have recently emerged as an ideal candidate for multi-messenger astronomy.  These large globular-shaped Galactic structures, extending up to $\sim 9$ kpc symmetrically out of the Galactic plane (see Fig.~\ref{FBgal}), were discovered \cite{Su:2010qj} in the gamma-ray data of the {\it Fermi} Large Area Telescope (LAT), which has since mapped them in details at energies $\sim 0.1$--400 GeV \cite{Fermi-LAT:2014sfa}. Complementary observations in radio \cite{FBradio} and X-ray \cite{FBXray} confirm multi-wavelength emission from the \fb. 
 
Both hadronic \cite{Crocker:2010dg,FBhadronic} and leptonic \cite{Su:2010qj,FBleptonic} mechanisms have been proposed for the origin of the \fb\ gamma rays.  For hadronic models, it has been pointed out that  a \n\ counterpart should exist \cite{Crocker:2010dg,Lunardini:2011br}, and might be responsible for a fraction of IceCube astrophysical neutrinos \cite{GalNu}. In particular up to five of the \ic\, astrophysical neutrinos of energy $\sim$ 100 TeV--1 PeV, that are spatially strongly correlated with the bubble geometry (see Fig.~\ref{FBgal}) \footnote{In refs.~\cite{ours} a distinction was made between \n\ events that correlate with the \fb\ strongly (central value of position within the bubbles' solid angle) and weakly (that overlap with the bubbles within the positional error). Here only the strongly correlated events are shown.}), are consistent with the hadronic flux model that reproduces the \fl\ gamma-ray data \cite{ours}.  Therefore, the \fb\ could be the first object to have been seen in both gamma rays and \ns.  The two observations, however, probe different parts of the \fb\ spectrum, and therefore their connection is only indirect.

An interesting development at the front of multi-messenger studies of the \fb\ is expected very soon, with the advent of the High Altitude Water Cherenkov detector (HAWC), which has started operations in Mexico this year \cite{Westerhoff:2014vka}. With a large  field of view ($\sim 2\pi$~sr), excellent positional resolution ($\sim 0.1^\circ$ at $\gtrsim 5$ TeV), and strong sensitivity in the $\sim 0.1$--100 TeV range, \hw\ will observe the \fb\ in great detail in a window of energy that partly overlaps with the  regimes already probed by the \fl\ and \ic. It will bridge the energy gap between them, thus leading to a more complete mapping of the \fb\ spectrum.  The interplay of the \fl,\ \hw\ and \ic\ data has the potential to disfavor or to fully establish the hadronic origin of the \fb, and, in the latter event, constrain the  spectrum of the primary proton/ion flux.  Such interdisciplinary study will be a good test bed for the development of new analysis techniques that could then be applied to other, less luminous, VHE \n\ and gamma ray sources \cite{sources} when \n\ telescopes reach the phase of high statistics data taking. 

The interplay between gamma ray and \n\ observations of the \fb\ is the focus of this paper.  
Here we study \hw's ability to detect VHE gamma rays from the \fb\ in the hadronic model, and present the first discussion of the potential of joint analyses of \hw\ and \ic\ data.

{\it Observing the \fb\ with HAWC and IceCube.\ ---} 
The FB are extended sources in the sky subtending a total solid angle $\Omega_{\rm FB}\simeq 0.808~{\rm sr}$ \cite{Su:2010qj}.  Depending on its field of view, location and time of the day, a detector on Earth will be able to observe only a fraction, $f_\Omega$, of this solid angle. Due to its privileged location near the South Pole,  \ic\ has $f_\Omega \simeq 1$ \cite{Lunardini:2011br},  meaning that it is equally sensitive to the \fb\ at all times of the day.   This is true for the High-Energy Starting Events (HESE) for which the $\nu N$ interaction vertex is within the detector volume~\cite{Aartsen:2013jdh}.

Instead, for \hw\ $f_\Omega$ is time-dependent. \hw\ is located in the Northern hemisphere, at longitude $97.3^\circ$ W and latitude $19.0^\circ$ N.  Its field of view has a half-opening angle $\delta \simeq 60^\circ$, meaning that it will be able to observe objects at angular distance $\theta < \delta$ from its zenith ($\cos\theta>\cos\delta = 0.6$).  Fig.~\ref{FBgal} shows the \fb\ in equatorial coordinates in comparison with the time-dependent region observed by \hw.  We see that the north bubble is observable entirely at least for 2-3 hours  in a day, while instead only a small portion of the south bubble is accessible to \hw.

For the purpose of obtaining event rates at \hw, we have calculated the daily-averaged fraction of the solid angle \cite{9780511861161}, $\langle f_{\Omega}(\theta_1,\theta_2) \rangle$,  that is subtended by the \fb\ and falls in angular bins defined by the values $\cos\theta = 0.6, 0.7, 0.8, 0.9, 1$. These bins  correspond   to the intervals for which the \hw\ effective are is reported \cite{Zaborov:2012jg}.  For each bin, in order of increasing $\cos\theta$, we find $\langle f_{\Omega}(\theta_1,\theta_2) \rangle = 4.5 \times 10^{-2}, ~3.5 \times 10^{-2}, ~4.1 \times 10^{-2} , ~1.0 \times 10^{-2}$.  On average, only a few percent of $\Omega_{\rm FB}$ falls within a certain zenith bin.  This plays the role of an efficiency factor in the calculation of the event rates.

\begin{figure}[htbp]
\centering
\includegraphics[width=3.3in]{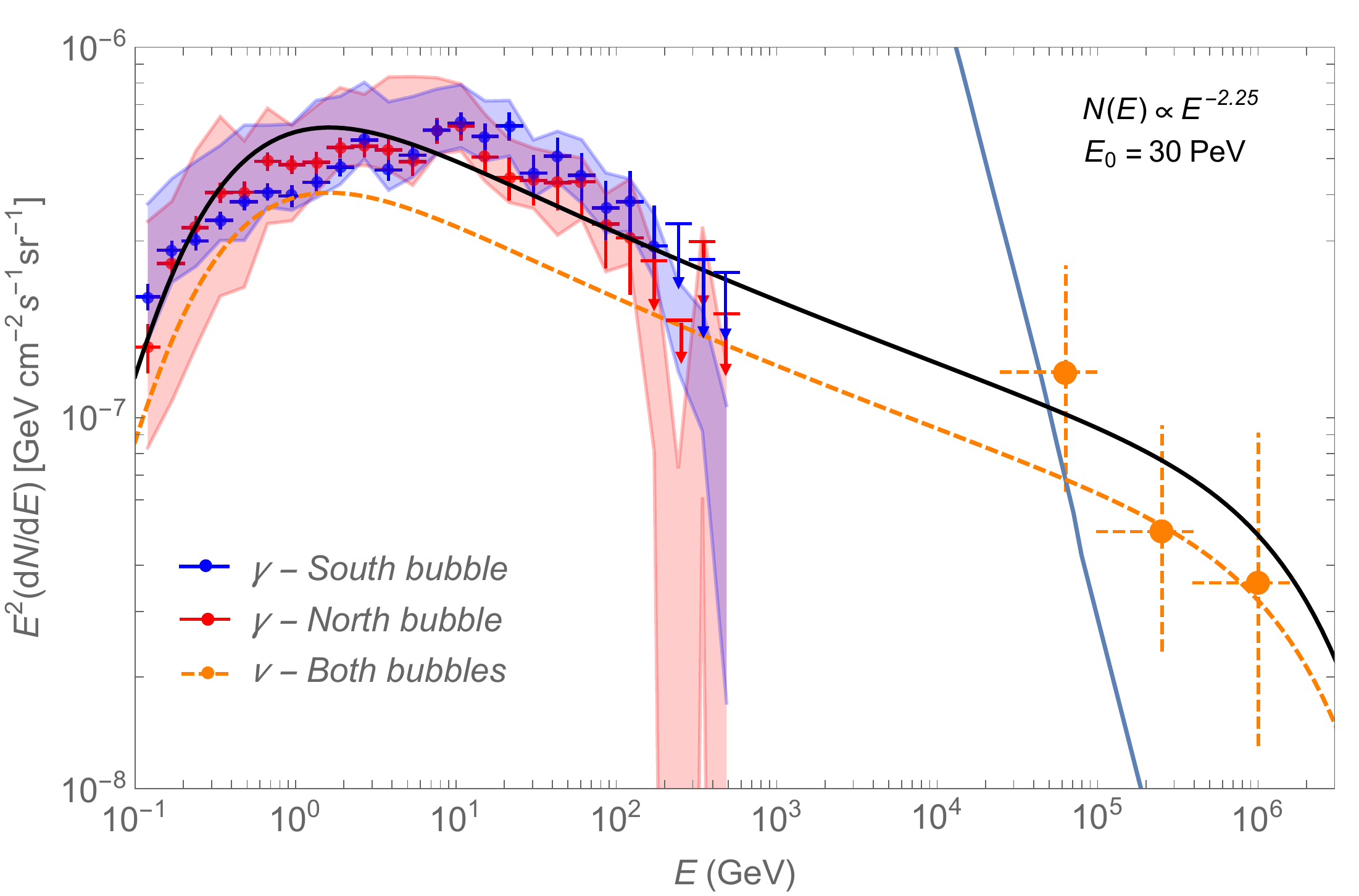}
\caption{The spectrum of the gamma-ray flux from the \fb\ (crosses), as measured by \fl\ \cite{Fermi-LAT:2014sfa}.  The shaded areas indicate the systematic errors.  The  solid curve is the corresponding prediction of a hadronic model (parameters in the inset).  The counterpart \n\ flux is shown too (dashed), and compared with the flux required by the \ic\ data (the three dots at high energies, with error bars), in the assumption that \fb\ are the sources of  the five \n\ events that are strongly correlated with them spatially \cite{ours}.  The \n\ errors assume Poisson statistics, following Pearson's $\chi^2$ intervals approach \cite{cdf1,cdf2}.  The atmospheric \n\ background \cite{Honda:2006qj}, averaged over $25^\circ-95^\circ$ zenith angle, is shown for comparison (steep solid line). }
\label{fluxes}
\end{figure}

We model the expected \n\ and gamma-ray fluxes in a simple hadronic model, with primary proton spectrum of the form $dN_p/dE \propto E^{-k} \exp(-E/E_0)$ and $pp$ interactions with dilute gas in the \fb, as described in details in ref.~\cite{Lunardini:2011br}. 
We focus on relatively hard spectra, with $k\simeq 2.2 - 2.3$ and $E_0 \sim 3-30$ PeV, that are required by the interpretation of the five \n\ data in fig. \ref{FBgal} as due to the \fb\ \cite{ours}.  In Fig.~\ref{fluxes}, the gamma ray and \n\ fluxes are shown for $k=2.25$, $E_0 = 30$ PeV.  Using a $\chi^2$ test, we checked that these spectra are a good fit of the \fl\ observation of both bubbles. In particular, $\chi^2/dof \approx 13/40$, including penalty for $\chi^2$ when the fit violates the upper limits (last four energy bins of the \fl\ data), which are treated as half-Gaussian.
 Of course, the \fl\ data alone are also compatible with a lower-energy cutoff in the proton spectrum, and slightly favor $E_0 \sim {\mathcal O}(10)$ TeV \cite{Fermi-LAT:2014sfa}. 
These values would require a different explanation, other than the \fb, for the \n\ data.

{\it Results:~VHE gamma-ray and neutrino event rates. ---}
As is typical of high energy cosmic-ray detectors, the response of \hw\ and \ic\ to a given flux $\Phi$ (of gamma rays and \ns, respectively) can be parameterized by an effective area, $A(E, \theta)$, which depends on the particle arrival direction and energy.  In the assumption that the \fb\ are uniformly bright in gamma rays (and hence in neutrinos in hadronic models) \cite{Su:2010qj,Lunardini:2011br,Fermi-LAT:2014sfa}, $\Phi$ is zenith-independent.  The number of events with primary particles in a certain zenith bin, $[\theta_1, \theta_2]$, and for an exposure time $T$, is given by \cite{Lunardini:2011br}
\beq
N &=& \int^{T}_0  dt  \int_ {
        \substack{\Sigma(t) \\ \theta_1 \leq \theta \leq \theta_2}
        } d\Omega   \int_{E_{th}}^\infty  dE \Phi(E) A(E, \theta) 
\nonumber \\
        &\simeq & T  \langle f_{\Omega}(\theta_1,\theta_2) \rangle \Omega_{FB}   \int_{E_{th}}^\infty  dE \Phi(E) \langle A(E)   \rangle_{\theta}~,~~~~~~
\label{nev2}
\eeq
where the zenith-averaged effective area (for each bin) is used as an approximation. Here $\Sigma(t)$ indicates that the integral in the solid angle is done over the region of the bubbles for which the condition on the zenith angle is satisfied.  An expression similar to eq.~(\ref{nev2}) also applies to the calculation of background rates in the two experiments, which is required to evaluate statistical significance of a signal from the \fb.

\begin{figure}[htbp]
\centering
\includegraphics[width=3.3in]{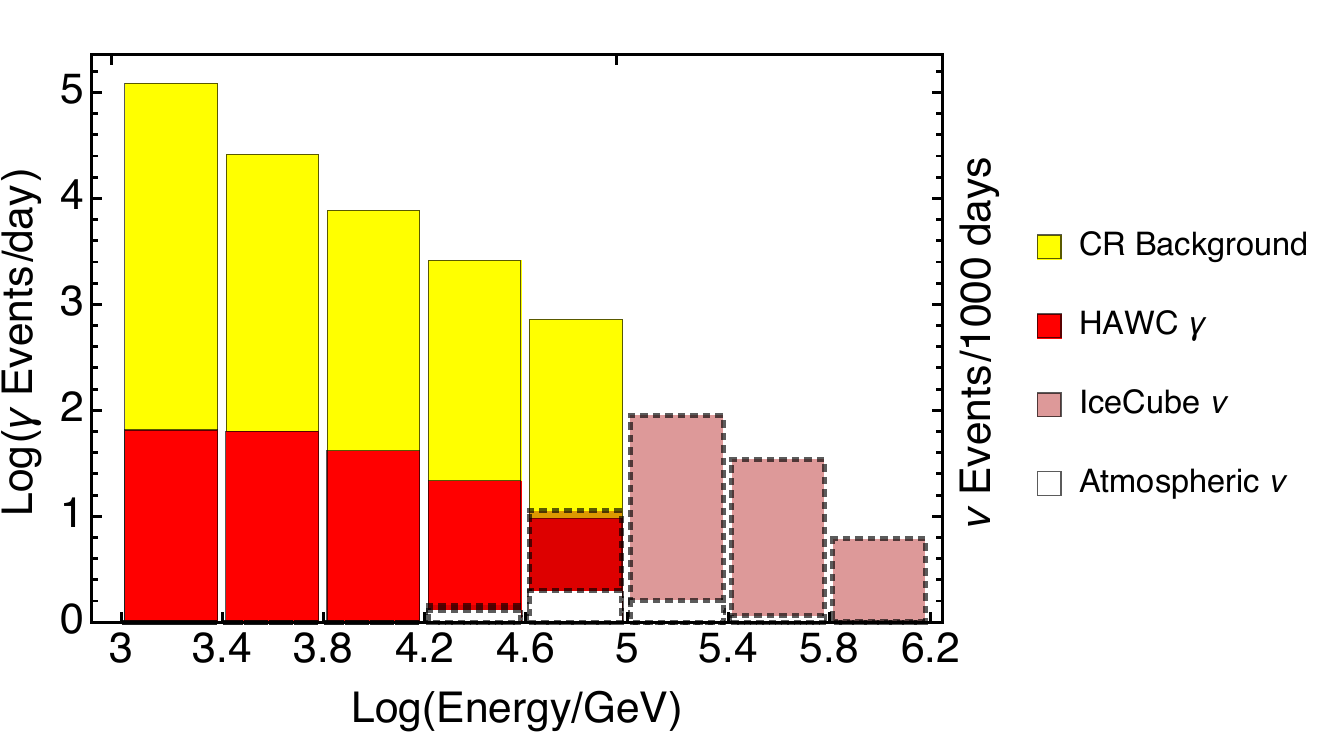}
\caption{{\it Solid histogram -} number of gamma ray events  (in logarithmic scale)   from the \fb\ at \hw, in bins of the gamma ray energy, for  1 day exposure (left vertical axis). The corresponding cosmic ray background is also shown. {\it Dashed histogram -} number of \n\ events (in linear scale) at \ic, in bins of \n\ energy. Here the exposure is $10^3$ days (right vertical axis), and the background is due to atmospheric \ns.  The primary flux parameters are the same as in Fig.~\ref{fluxes}. }
\label{nevents}
\end{figure}

For \n\ event rate calculation in \ic, we adopt the same method as in \cite{ours}, using the averaged effective area as in ref.~\cite{Aartsen:2013jdh} for each \n\ flavor. The main background here is due to the flux of atmospheric \ns\ \cite{Honda:2006qj}, shown in Fig.~\ref{fluxes}.  The \fb\ and atmospheric \n\ flux models well-reproduce the 5 cascade events that are strongly-correlated with the \fb\ spatially, over a 1000 day \ic\ live time (see Fig.~\ref{nevents}).

For the VHE gamma-ray event rate at \hw, both for signal and background,
we use the effective area in ref.~\cite{Zaborov:2012jg}.  The absorption of gamma rays due to electron-positron pair production with photons from starlight or cosmic microwave background is negligible in the \hw\ energy range. The dominant background in \hw\ is due to cosmic rays, mostly protons and helium nuclei, whose fluxes are $\sim 4-5$ orders of magnitude larger than the flux of the \fb\ gamma rays.  We use the hadron rejection efficiency for \hw, which is $5\times 10^{-3}$ at energies above 10 TeV, from ref.~\cite{Abeysekara:2013tza}.  Another background is diffuse gamma-ray emissions, which have been well measured by \fl\ \cite{FermiLAT:2012aa}.   To estimate the gamma-ray background in different sky regions, we adopt the spectra for the $^SS^Z4^R20^T150^C5$ model (extrapolated to the energies of interest here) in ref.~\cite{FermiLAT:2012aa} for the inner Galaxy region (see also ref.~\cite{Gaggero:2014xla}  for gamma-ray background in this region), high and low intermediate latitude region as shown in Fig.~\ref{FBgal}.  The cosmic electron background is negligible.

Figure~\ref{nevents} shows the signal and background daily event rates in \hw\ from the \fb\ region. With several signal events per day, \hw\ will rapidly accumulate a high statistics data sample. Although the signal from the \fb\ is $\sim 2-3$ orders of magnitude smaller than the background, the high statistics will allow to establish a high significance, given by the number of Gaussian standard deviations with respect to background as $\sigma = \sqrt{\sum_i \sigma_i^2} = \sqrt{T} \sqrt{\sum_i S_i^2/(S_i + B_i)}$.  Here $S_i$ and $B_i$ are \fb\ signal and background (total of cosmic-ray and diffuse gamma-ray fluxes) event rates, respectively, in each energy bin $i$.  We find that, for the total of all energy and zenith bins, $\sigma>3$ ($\sigma>5$) already after 10 days (35 days) of running time.  After a year of operation, a significance of at least 5$\sigma$ will be reached in each zenith bin separately.

{\it Discussion:~HAWC-IceCube complementarity. ---}
We have found that \hw\ has an excellent potential to observe the \fb\ with high significance within a relatively short time scale.  Depending on the parameters, this signal could be consistent with the hypothesis that the bubbles might be the (hadronic) source of the \ic\ \n\ events that spatially correlate with them. Let us outline below what can be learned from \hw\ and \ic, in combination and individually, on the \fb.

\begin{figure}[htbp]
\centering
\includegraphics[width=3.3in]{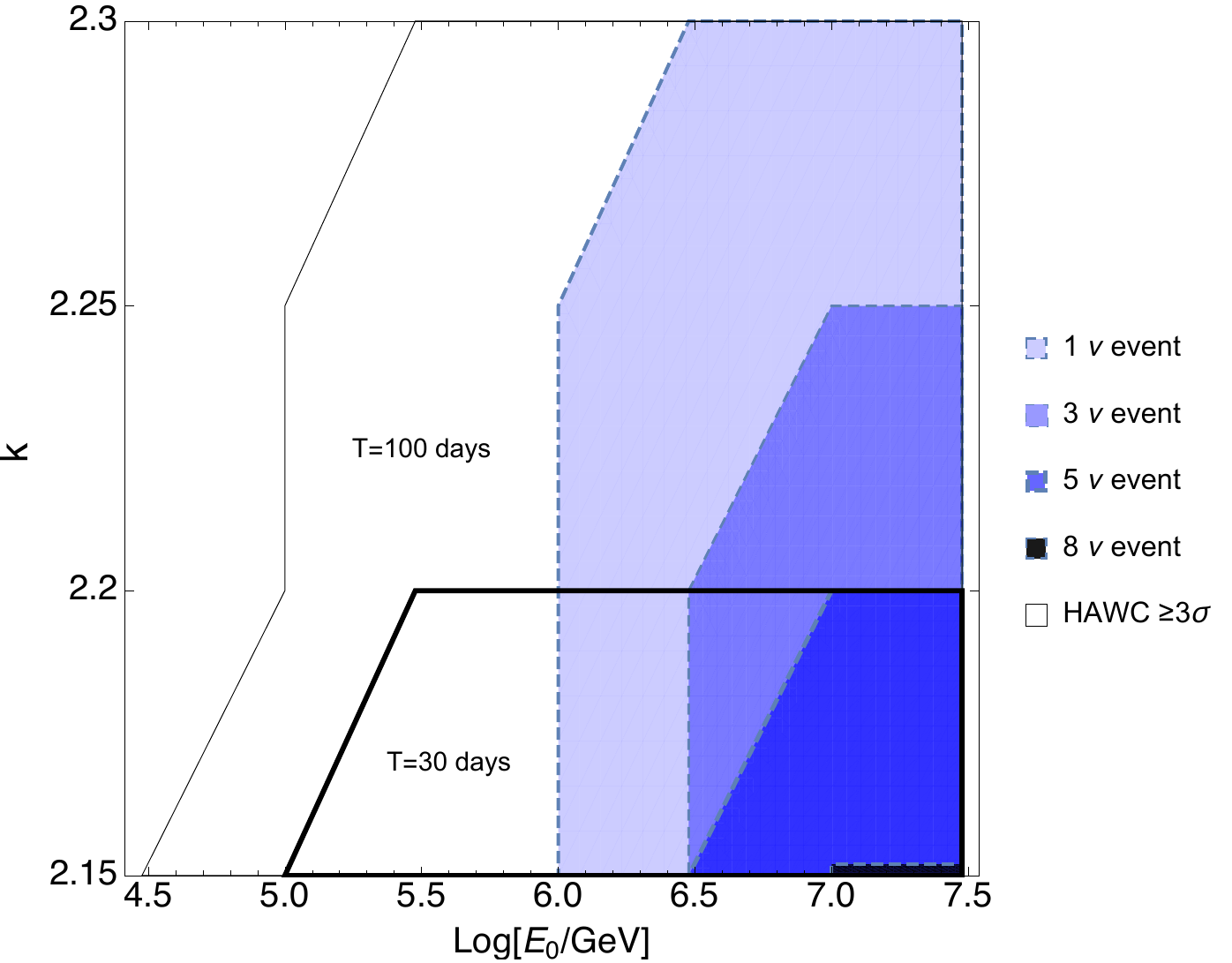}
\caption{Solid lines:  contours in the space of the flux parameters, $k$ and $E_0$, corresponding to more than $3\sigma $ significance at \hw\ (for two different running times, see legend). Dashed, shaded regions: isocontours of  numbers of events at \ic, for 1000 days running time.  These numbers of events have been calculated at discrete points on a grid with step 0.5 (0.05) on the horizontal (vertical) axis. }
\label{contours}
\end{figure}

Figure~\ref{contours} elaborates on the multi-messenger connection between the \ic\ and \hw, showing the regions of the parameter space ($k=2.15 - 2.30$ and $E_0= 10^{4.5 } - 10^{7.5}$ GeV) that correspond to a given signal at the two detectors (significance for the \hw\ and number of events for \ic). 
 For each pair ($E_0$, $k$), the flux normalization is determined by the low-energy 
Fermi-LAT data for the $k$ and $E_0$ values, similar to the fit shown in Fig.~\ref{fluxes}.
It can be seen that, while \hw\ can probe the entire space with high statistical significance within a year or so of operation, \ic\ is insensitive to a \n\ flux with a spectral cutoff in the primary proton energy $E_0 \lesssim 1$ PeV, as expected, considering the higher threshold of \ic. Therefore, the different combinations of possible outcomes (detection or exclusion) at the two observatories will be informative of the spectral parameters.

\ic\ has already observed 5 \n\ events that are correlated in position with the \fb\ (see Fig.~\ref{FBgal}).  With a few more years of operation, this hint could become a statistically significant observation of the \fb.  This observation would confirm the hadronic hypothesis for the bubbles, and indicate a primary proton population with a relatively hard spectrum, $k\simeq 2.15 - 2.25$, and high energy cutoff, $E_0 \simeq 3 - 30$ PeV.  It may provide the best information on the highest energy tail of the \fb\ spectrum, $E \gta 0.1 $ PeV, which is beyond the sensitivity of \fl\ and \hw.  Furthermore, \ic\ will observe both bubbles, therefore testing the degree of symmetry between them, in a way that complements the lower energy information from \fl.  A negative result at \ic\ will be compatible with either a hadronic scenario with sub-PeV cutoff, or a primary leptonic origin of the \fb.

If \ic\ observes the \fb, and therefore the primary spectrum extends above PeV, a high statistics observation in gamma rays is expected at \hw\ within about a year of operation.  This telescope is mainly sensitive to the North Bubble and will produce a detailed intensity map of it, at energies higher than those probed so far by the \fl.  This map will test a number of features seen by \fl, like the uniform projected intensity of the bubbles \cite{Su:2010qj,Fermi-LAT:2014sfa}.  Due to its lower energy threshold, \hw\ could see a signal even for a negative result at \ic. In particular, thanks to its good energy resolution, \hw\ could reveal the presence of a spectral cutoff at the $\sim 0.1-100$ TeV energy window.

{\it Outlook. ---} Looking at the longer term future, a larger overlap between the energy windows of the two experiments would be desirable, especially to fully exclude the hadronic model in the absence of a signal at \ic, accompanied by a clear detection at \hw\ in the same energy range.  This larger overlap could be achieved by lowering the energy threshold of the \ic\ analysis.  It has to be considered that atmospheric \n\ backgrounds at \ic\ become quickly overwhelming when lowering the threshold, therefore the potential of such exercise is unclear at this time. 

Besides a direct comparison of the \ic\ and \hw\ data, the indirect  influence of one experiment over the other will be important as well.  Indeed, each data set will lead to improved theoretical models of the \fb, which in turn will result in more precise predictions for the next generation of experiments. In particular, the ratio of gamma ray and \n\ fluxes may be constrained and compared with predictions, to help discriminate between different versions of the hadronic model and possibly test \n\ and/or gamma ray propagation effects over galactic scales of distance. Similar multi-messenger studies for a point source or diffuse emission at the Galactic Center region can be done as well~\cite{GCMultiMessenger}.

Eventually, the interdisciplinary research on the \fb\ with \ns\ and gamma rays will include more participants. One of them will be KM3NeT, the multi-km$^3$ \n\ detector currently planned in the Mediterranean sea \cite{Adrian-Martinez:2012qpa}. It's current predecessor ANTARES has already searched for a signal from the \fb\ direction, resulting in a flux upper limit \cite{Adrian-Martinez:2013xda}.  Relative to \ic,  KM3NeT will have a larger effective area ($\sim 6$~km$^3$ instrumented volume) and may detect the \fb\ in about one year of its operation \cite{Adrian-Martinez:2012qpa}. 
Finally, \ic\ itself might be able to probe the highest energy end ($\gtrsim 1$ PeV) of the \fb\ gamma-ray spectrum using the IceTop surface detector array \cite{Aartsen:2012gka}.  Prospects for PeV gamma-ray detection have also been considered for the upcoming ground-based detectors LHAASO (Large High Altitude Air Shower Observatory)~\cite{Cui:2014bda} and HiSCORE (Hundred*i Square-km Cosmic ORigin Explorer)~\cite{Tluczykont:2014cva}.

We conclude that, by extending the gamma ray mapping of the \fb\ to 100 TeV, \hw\ will be an important complement to \n\ searches of these objects. With \ic\ and other  \n\ observatories, it will further advance the new field of multi-messenger astronomy at very high energy.

{\it Acknowledgements. ---} We thank Paolo Desiati, Brenda Dingus, Anna Franckowiak, Thomas Gaisser, Javier Gonzalez, Kohta Murase and Juan Jos\'e Rey Hern\'andez for helpful discussion.  Work of S.R.\ was supported in part by the National Research Foundation (South Africa) Grants No.\ 87823 (CPRR) and No.\ 91802 (Blue Skies). C.L.\ acknowledges the National Science Foundation grant number PHY-1205745.



\end{document}